\documentstyle[12pt]{article}
\def \a{\alpha}
\def \b{\beta}
\def \l{\lambda}
\def \d{\delta}
\def \p{\partial}
\def \dg{\dagger}
\def \e{\epsilon}
\def \s{\sigma}
\def \t{\theta}

\def \Pm{\pmatrix}
\def \be{\begin{eqnarray}}
\def \ee{\end{eqnarray}}
\def \no{\nonumber}

\begin{document}

\centerline{\large{\bf Solitons in a gauged Landau-Lifshitz model}}

\vskip 0.5in

\centerline{B. Chakraborty\footnote{e-mail: biswajit@boson.bose.res.in}
and A. S. Majumdar\footnote{e-mail: archan@boson.bose.res.in}}

\vskip 0.3in

\centerline{S. N. Bose National Centre for Basic Sciences}

\centerline{Block-JD, Sector-III, Salt Lake, Calcutta-700091, India.}

\vskip 2.0in

We study the gauged Landau-Lifshitz model in which 
a suitably chosen triplet of background
scalar fields is included. It is shown that the model admits
solitonic configurations.

\pagebreak

Ferromagnetic systems in two spatial dimensions can be described
by the Landau-Lifshitz (LL) model [1] in the long wavelength limit.
The discrete version of this model is equivalent to the Heisenberg
model. This model involves time derivatives of the first order only, and is
actually a nonrelativistic sigma model. The LL model is also equivalent
to the nonrelativistic $CP^1$ model~[2]. The relativistic counterpart,
i.e., the standard $O(3)$ nonlinear sigma model (NLSM), on the contrary,
describes antiferromagnetic systems~[3].

It was shown by  Nardelli~[4] that the gauged NLSM containing an $SO(3)$
Chern-Simons (CS) term admits a new kind of
topological soliton which can not  be chracterised by $\Pi_2$-the
second homotopy group, unlike NLSM. Later, Cho and Kimm~[5] generalized
the result for arbitrary $CP^{N-1}$, where the global $SU(N)$ group is
gauged and a corresponding CS term is added. Thus the question naturally
arises, whether the gauged nonrelativistic $CP^1$ model also admits
topological solitons similar to its relativistic counterpart. We try to
analyse this issue in the present letter.

To this end, we consider the model given by
\be
{\cal L} = {i\over 2}\biggl[(D_0Z)^{\dg}Z - Z^{\dg}(D_0Z)\biggr] 
- |D_iZ|^2 - a_0(Z^{\dg}Z-1) \no \\
+ \t \e^{\mu\nu\l}(A^a_{\mu}\p_{\nu}A^a_{\l} + {g\over 3}\e^{abc}
A^a_{\mu}A^b_{\nu}A^c_{\l}) + {g\over 2}A^a_0\phi^a
\ee
Here we have introduced  a triplet of real scalar (background) fields
$\phi^a$, for reasons that will become clear later, and 
$Z= \Pm{z_1 \cr z_2}$ is a complex doublet satisfying
$Z^{\dg}Z = 1$. $\t$ represents the CS parameter. The covariant derivatives 
are given by
\be
D_0 = \p_0 -igA^a_0T^a \no \\
D_i = \p_i -ia_i -igA^a_iT^a
\ee
with $T^a = \s^a/2$. Here $a_i$ and $A^a_i$ represent the $U(1)$
and $SU(2)$ gauge fields respectively. Note that since we are
considering a non-relativistic model, we have the freedom to introduce
different temporal and spatial 
 ``covariant'' derivatives in (2) without violating any
principles.

The Legendre transformed Hamiltonian density which is given by
\be
{\cal H} = |D_iZ|^2 + a_0(Z^{\dg}Z-1) + A^a_0\biggl({g\over 2} (M^a-\phi^a)
-2\t B^a\biggr)
\ee
where $B^a  = (\p_1A^a_2 - \p_2A^a_1 + g\e^{abc}A^b_1A^c_2)$ is the non-abelian
$SU(2)$ magnetic field.

Clearly, $a_0$ and $A^a_0$ are the Lagrange multipliers enforcing the
constraints
\be
G_1 = (Z^{\dg}Z - 1) \approx 0 \no \\
G_2^a = {g\over 2}(M^a-\phi^a) - 2\t B^a \approx 0
\ee
As this nonrelativistic model is first order in time derivative, the
symplectic structure can readily be obtained by using the Faddeev-Jackiw~[6]
method to yield the following brackets:
\be
\{z_{\a}(x),z^*_{\b}(y)\} = i\d_{\a\b}\d(x-y) \no \\
\{A^a_i(x),A^b_j(y)\} = {\e_{ij} \over 2\t}\d^{ab}\d(x-y)
\ee
$G_1$ and $G^2_a$ are the first class Gauss constraints of this model, and
it can be shown that $G_1$ and $G^2_a$ generate the appropriate $U(1)$
and $SU(2)$ transformations respectively, i.e.,
\be
\d Z(x) = \int d^2y f(y) \{Z(x),G_1(y)\} = if(x)Z(x) \no \\
\d Z(x) = \int d^2y f^a(y) \{Z(x),G^a_2(y)\} = {ig\over 2}f^a(\s^aZ) \no \\
\d A^a_{\mu}(x) = \int d^2 y f(y) \{A^a_{\mu}(x), G_1(y)\} = 0 \no \\
\d A^a_i(x) = \int d^2y f^b(y) \{A^a_i(x), G^b_2(y)\} = \p_if^a(x) + g\e^{abc}
A^b_if^c
\ee

The momentum variable $\pi_i$ conjugate to $a_i$ vanishes,
\be
\pi_i = {\d {\cal L} \over \d \dot{a}_i} = 0
\ee
Preservation of this primary constraint in time yeilds the secondary
constraint
\be
a_i \approx -i Z^{\dg} \p_i Z - {g\over 2} A^a_iM^a
\ee
where 
\be
M^a = Z^{\dg}\s^a Z
\ee
 is a unit vector obtained by using the Hopf
map. Clearly, (7) and (8) form a pair of second class
constraints, and therefore are `strongly' implemented by the Dirac bracket
\be
\{a_i, \pi_j\} = 0
\ee
With this, $a_i$
ceases to be an independent degree of freedom. It may be mentioned here
that the same thing happens to the $U(1)$ gauge field for the case of
the gauged $CP^1$ model coupled to the CS term in its relativistic
avatar which was studied in [7].

One can show that $G^a_2$ satisfy an algebra isomorphic to the $SU(2)$
algebra.
\be
\{G^a_2(x), G_2^b(y)\} \approx 2\e^{abc}G^c_2(x)\d(x-y)
\ee
Following the group theoretical arguments, as in [7], one can show that
the following relations between the Gauss constraints are expected. 
\be
gG_1 = M^aG^a_2
\ee
This can be verified using
$M^aM^a = (Z^{\dg}Z)^2 \approx 1$. This shows that $G_1$ is not an
independent constraint,
and, this model therefore, has only $SU(2)$
gauge invariance. This is again similar to the case of the relativistic
version of the gauged $CP^1$ model coupled to the Hopf term [7].

We are now equipped to address the question of solitonic configurations [8]
in this model. 
To investigate the existence of solitons, consider the energy functional
obtained from the Hamiltonian density (3) (on the constraint surface(4)),
which is given by
\be
E = \int d^2x (D_iZ)^{\dg}(D_iZ)
\ee
which can be rewritten as
\be
E = \int d^2x |(D_1 \pm iD_2)Z|^2 \pm 2\pi N
\ee
where
\be
N = {1\over 2\pi i}\int d^2x \e_{ij} (D_iZ)^{\dg}(D_jZ)
\ee
In order for solitons to exist, the corresponding static configurations must
satisfy the saturation condition
\be
(D_1 \pm iD_2)Z = 0
\ee
and $N$ should be given by some number which may have a topological 
origin. Note that $N$(15) is SU(2) gauge invariant, and so we can 
evaluate $N$ in any gauge of our choice. At this stage we rewrite $N$ as
\be
N= {1\over 4\pi} \int d^2x \e_{ij} \biggl[-2 i \p_iZ^{\dg}\p_jZ
+ gA^a_i(\p_jM^a + {g\over 2} \e^{abc} A^b_j M^c)\biggr]
\ee
By making use of the local $SU(2)$ gauge invariance
of the model, we can go to a configuration where $Z(x) =$ constant. 
Correspondingly, $M^a$'s are also constant, so that quantities like
$\p_jZ$ and $\p_jM^a$ vanish, and $N$ (17) reduces to  
\be
N = {g^2 \over 4\pi} \int d^2x \e^{abc} M^a A^b_1 A^c_2
\ee

On the other hand, the invariance under $SU(2)$ (global) transformation  
of the model (1) implies
the existence of the following triplet of conserved $SU(2)$ charges
\be
Q^a = 2\t \int d^2x (\p_1A^a_2 - \p_2 A^a_1)
\ee
as follows from Noether's theorem. In terms of the corresponding charge
density $J^{0a} = 2\t (\e_{ij}\p_i A^a_j)$, $N$ becomes
\be
N = {g\over 4\pi} M^a \int d^2x \biggl(B^a - {J^{0a} \over 2\t}\biggr)
\ee
Making use of the Gauss constraint $G_2$ (4), one can
write
\be
N = {g^2 \over 16\pi \t} \int d^2x (1 - M^a \phi^a) - {g\over 8\pi\t}M^aQ^a
\ee
Now if the background field $\phi^a$ satisfies the condition 
\be
M^a\phi^a = 1
\ee
 then using (21) and (4) one gets
\be
N = - {g\over 8\pi\t}M^aQ^a \\
M^aB^a = 0 
\ee
Note that all $(M^a\phi^a)$, $(M^aB^a)$, and $(M^aQ^a)$ are $SU(2)$ scalars.

At this point we make a particular 
gauge choice\footnote{Although the presence of the CS term allows for
only those gauge transformations which tend to a constant at infinity,
$E$(13) and $N$(15) are invariant under arbitrary gauge transformations.}
 $Z = \Pm{0 \cr 1}$. 
Correspondingly, $M^a = - \d^{a3}$, and
\be
B^3 = \p_1 A^3_2 - \p_2 A^3_1 + g \e^{\a\b}A^{\a}_1A^{\b}_2 = 0
\ee
($\a,\b = 1,2$). With this $Q^3$ (19) can be reexpressed as
\be
Q^3 = -2 \t g \int d^2x (A^1_1A^2_2 - A^1_2A^2_1)
\ee
(Just to remind the reader, the subscripts and superscripts of $A$ stand
for spatial and group indices respectively.) Note that in this gauge, one
has only a surviving $U(1)$ symmetry. This corresponds to an $SO(2)$
rotation around the $M^3$ axis. $Q^3$ is the corresponding conserved
(Noether) charge.
Making use of (2) and (8), one of the saturation conditions (16), 
(i.e., $(D_1 + iD_2)Z = 0$) in the gauge $Z = \Pm{ 0 \cr 1}$, can be shown
to yield
\be
A^1_1 = - A^2_2 \no \\
A^1_2 = A^2_1
\ee
Using (26) and (27) one finds 
\be
Q^3 = 2\t g \int d^2x \bigl((A^1_1)^2 + (A^1_2)^2\bigr) =
2\t g \int d^2x \bigl((A^2_2)^2 + (A^2_1)^2\bigr) > 0 
\ee
and consequently (using (23)),
\be
N = {g\over 8\pi\t}Q^3 > 0
\ee
and the minimum value of energy is
\be
E_{(min)} = 2\pi N = {g\over 4\t} Q^3 =
{g^2 \over 2}\int d^2x \biggl((A^1_1)^2 + (A^1_2)^2\biggr) > 0
\ee
One can easily check that the other saturation condition
$((D_1 - iD_2)Z=0)$ yields
\be
A^1_1 = A^2_2 \no \\
A^1_2 = - A^2_1
\ee
and $Q^3$ (26) becomes
\be
Q^3 = -2\t g \int d^2x \biggl[(A^1_1)^2 + (A^1_2)^2\biggl] < 0
\ee
With this, $N$ is given as
\be
N = {g \over 8\pi\t}Q^3 < 0
\ee
But now, $E_{min}$ is given by
\be
E_{min} = -2\pi N = - {g \over 4\t}Q^3 =
{g^2 \over 2}\int d^2x \biggl[(A^1_1)^2 + (A^1_2)^2\biggl] > 0
\ee
Thus, either of the saturation conditions (16) yields the same
(positive definite) value for $E_{min}$ as desired. The only difference is
that the two conditions in (16) correspond to the positive and negative
values for the number $N$.

Thus we must have either of the
saturation conditions corresponding to the configuration $Z = \Pm{ 0 \cr 1}$.
Also note that for $E_{(min)}$ to be finite, $A^1$ and $A^2$ must vanish
asymptotically. Thus asymptotically, $B^3$ (25) becomes a $U(1)$ magnetic
field with $A^3_i$ as the abelian gauge field. The points at infinity can be
identified, so that the two-dimensional plane gets effectively compactified
to $S^2$. From (19), it then follows that $Q^3$ represents a topological
index, which is nothing but the first Chern class. Clearly, $E_{min}$ 
((30) and (34)) is
given by a topological index. Thus these field configurations satisfying
either of the saturation conditions (27) or (31) correspond to topological
solitons, with positive or negative ``winding numbers'' ($N$). The same
holds for all other configurations obtained by gauge transformations. For
example, for any other configuration with $Z(x) =$ constant, 
the $E_{min}$ will be given
by an appropriate $Q^a$ obtained by making a suitable 
$SO(3)$ rotation corresponding
to the $SU(2)$ transformation required to get $Z = $ constant configuration from
$Z = \Pm{0 \cr 1}$. But the value of $E_{min}$ will remain the same. It
is important to note that we had to choose a convenient gauge here to
identify $E_{min}$ with a topological index (up to a constant). It is 
rather nontrivial to make this identification in an arbitrary gauge.

Note further, that if in adition, $\phi^a = M^a$, then $B^a=0$ (4)
and $A_i^a$ are pure gauges. In this case we can go to a gauge
where $A^a_i=0$ and $D_i|_{A_i=0} = {\cal D}_i$ (the covariant 
derivative operator of the pure $CP^1$ model). Here also $N$ corresponds
to the topological index (Chern class) and we have a topological 
soliton.

Finally, note that for $\phi^a = 0$, the first term in $N$ (21) diverges,
and there does not exist any solitonic configuration. It was thus necessary
to introduce a triplet of background scalar fields $\phi^a$ in (1),
satisfying (22) in order to obtain solitonic configurations. This is in
contrast to the case of the relativistic model~[5] and its reduced
phase space version~[7].

\pagebreak

REFERENCES

\vskip 0.5in

\begin{description}

\item[[1]] L. Landau and E. Lifshitz, Phys. A (Soviet Union) {\bf 8}
(1935) 153; A. Kosevich, B. Ivanov and A. Kovaler, Phys. Rep. {\bf 194}
(1990) 117.
\item[[2]] E. Flotaror, Phys. Lett. B {\bf 279} (1992) 117; R. Banerjee
and B. Chakraborty, Nucl. Phys. B {\bf 449} (1995) 317.
\item[[3]] E. Fradkin, ``Field Theories in Condensed Matter Systems''
(Addison Wesley, Redwood City, CA, 1991).
\item[[4]] G. Nardelli, Phys. Rev. Lett. {\bf 73} (1994) 2524.
\item[[5]] Y. Cho and K. Kimm, Phys. Rev. D {\bf 52} (1995) 7325.
\item[[6]] L. Faddeev and R. Jackiw, Phys. Rev. Lett. {\bf 60} (1988) 1692.
\item[[7]] B. Chakraborty and A. S. Majumdar, Phys. Rev. D {\bf 58} (1998) 125024.
\item[[8]] R. Rajaraman, ``Solitons and Instantons'' (North Holland, 1982);
C. Rebbi and G. Soliani, ``Solitons and Particles'', (World Scientific,
Singapore, 1984); V. A. Novikov, M. A. Shifman, A. I. Vainshtein and
V. I. Zhakharov, Phys. Rep. {\bf 116} (1984) 103.

\end{description}

\end{document}